# Quantum Dynamical Relativity and the Nature of Minkowski Spacetime

Allan Walstad

Physics Department
University of Pittsburgh at Johnstown
Johnstown, PA 15904 USA
awalstad@pitt.edu

ABSTRACT

Einstein's special theory of relativity starts with assumptions about how observations conducted in relatively moving inertial frames must compare. From these assumptions, conclusions can be drawn regarding the laws of physics in any one frame. Can we do the reverse, that is, build the correct physics entirely from considerations within a single frame and infer therefrom the results of measurements in moving frames? Can we, in this project, supply the "elementary foundations" that Einstein found lacking in his own theory, thereby converting special relativity from (in Einstein's terminology) a "principle" theory to a "constructive" one? Expanding on the approach suggested by John S. Bell, this paper demonstrates that we can progress toward those goals via the introduction of elementary quantum concepts, particularly the idea of matter waves (deBroglie waves) that was unknown to Einstein in 1905. This quantum dynamical approach offers to resolve a recent dispute in philosophy of physics regarding the Minkowski geometry of spacetime.

## 1 Introduction

John Bell's well-known collection of essays *Speakable and Unspeakable in Quantum Mechanics* contains an entry that appears rather out of place in a book on quantum philosophy. "How to Teach Special Relativity" (Bell [1982]) advocates what Bell calls a "Lorentzian pedagogy," based on the efforts of Lorentz, Fitzgerald, Poincare and others to resolve the same theoretical and observational puzzles about electromagnetism and light that eventually led Einstein to his relativity postulates. Where Einstein invoked observers in relatively moving inertial frames of reference, Bell argues for an approach based on the application of electromagnetism and classical (non-quantum) dynamics in a single frame. In this approach, it is dynamics that answers our questions about the results of measurements made by moving observers, rather than assertions about moving observers constraining the dynamics.

For Einstein's predecessors, the "single frame" would have been defined by the ether, the all-pervasive elastic medium that was presumed necessary to support the propagation of electromagnetic waves. Attempts to reconcile ether theory with the results of experiments, including the Michelson-Morley null result, had already led to proposal of ideas such as length contraction and time dilation that we now associate with relativity theory. It was known that Maxwell's equations are covariant under a Lorentz transformation, and it was strongly suspected that absolute motion (i.e., motion relative to the ether) would prove completely unobservable.[1]

A dynamical approach does not require the existence of an ether or preferred frame of rest. For the analysis to get underway, it is only necessary that the laws of physics be established in one reference frame. If we can show from the resulting dynamics that the same laws of physics and same speed of light are recorded by measurement devices moving with constant velocity relative to that frame, then we will have demonstrated what Einstein had only assumed. With his relativity postulates, Einstein was able to circumvent considerations regarding the properties of the ether and the internal structure of physical bodies while straightforwardly deriving numerous important results.[2] Nevertheless, he acknowledged that something significant by way of understanding had been left behind.

Einstein regarded the relativity postulates as playing a role similar to that of the laws of thermodynamics. From the latter can be deduced a wide range of valuable results, yet we would not consider ourselves truly to understand thermal phenomena until we can trace their origin to the underlying physics, such as the random motions of molecules in an ideal gas: that is, to the theory of statistical mechanics. Special relativity and thermodynamics are "principle" theories, according to Einstein, while statistical mechanics would be a "constructive" theory, having as yet no satisfactory relativistic counterpart.[3] It seems clear that Einstein saw the constructive approach, where available, as more satisfactory. In an essay (Einstein [1919]) for the London Times, for example, he offered the following statement: "When we say we have succeeded in understanding a group of natural processes, we invariably mean that a constructive theory has been found which covers the processes in question." In a letter (Einstein [1908]) to Arnold Sommerfeld he had been more emphatic: "It seems to me too that a physical theory can be satisfactory only when it builds up its structures from *elementary* foundations. The theory of relativity is not more conclusively and absolutely satisfactory than, for example, classical thermodynamics was before Boltzmann had interpreted entropy as probability. If the Michelson-Morley experiment had not put us in the worst predicament, no one would have perceived the relativity theory as a (half) salvation." [Emphasis in the original.]

Can the "elementary foundations" that Einstein found lacking in special relativity be supplied? Bell [1982] is an effort in that direction.[4] Bell investigates what would happen to the orbit of an electron in a hydrogen atom if the proton were very gently accelerated to high speed. His model of the atom is purely classical, with the electron initially following a circular orbit in the xz plane, the proton being accelerated from rest along the z axis to speed v. Numerically integrating the equations of motion for the trajectory of the electron in the electromagnetic field of the moving proton, one finds that its orbit is flattened by a factor of $1/\gamma$ along the z direction, where $\gamma = (1 - v^2/c^2)^{-1/2}$ is the familiar Lorentz factor and c is the speed of light. Thus, length contraction, interpreted relativistically as a kinematic or geometric effect, becomes reinterpreted as a dynamical one. Bell also finds that the orbital period is lengthened by a factor of $\gamma$, which we recognize as the relativistic phenomenon of time dilation.

In deriving these results, Bell makes use of what we call the "relativistic momentum formula" $p = m\gamma v$. This he attributes to Lorentz, but without a citation. In Lorentz [1904], equation 28 is in fact the correct expression, upon substitution for various parameters from elsewhere in the paper. Lorentz's derivation itself is long and arduous and applies only to the electron as a purely electromagnetic phenomenon.

## 2 "Relativistic" Particle Dynamics Without Relativity

Bell was on firmer ground than he seems to have realized. The correct expression for momentum of a particle traveling at high speed was already implicit in physics going back to Maxwell, easily derived and not limited to the special case of an electron. A directed pulse of light with energy E has momentum $p = E/c$. From this, as Einstein [1906] showed in an early thought experiment—quite independently of the relativity postulates—it follows that energy must be associated with mass via the familiar $E = mc^2$. (For a textbook discussion of "Einstein's box" see French [1968], pp. 16-29.) With this result in hand, a slightly modified version of the same thought experiment, not employed by Einstein or found elsewhere, establishes the relation $p = (m + K/c^2)v$ among the momentum, (rest) mass, kinetic energy, and velocity of a material particle. From this equation it is easy to derive the momentum formula $p = m\gamma v$. Details follow.

Einstein's thought experiment involves a box of length L and mass M, as illustrated in Figure 1.

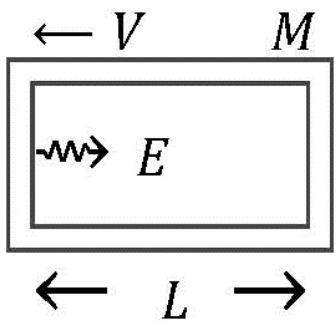

Figure 1: Einstein's box.

A pulse of light having energy E is emitted from the left end and absorbed at the right. When the light is emitted, the box must recoil with momentum $p = E/c$, hence with speed $V = E/Mc$ (which we are taking to be slow enough that there is no question about the accuracy of the Newtonian expression for momentum). The box is brought to rest when the light is absorbed. Nevertheless, during the light travel time $t = L/c$, the box moved to the left a distance δs = Vt = EL/Mc². Thus it would appear that the center of mass of an isolated system has shifted spontaneously.  (Indeed, in Einstein's discussion, if the energy can be returned by means other than light back to the left side, the process can be repeated indefinitely, flagrantly violating conservation of momentum.)  This result can be avoided if the energy transferred is associated with a mass δm = E/c².  In that case we will have M δs = L δm, which is the condition that the center of mass not move. (Einstein's argument generated a small literature of tweaks to improve its rigor.  See Antippa [1976] for a review.)  Direct observation of mass-energy equivalence came with the development of sufficiently accurate mass spectrometers in the 1930s.

We consider now an apparently novel extension of the thought experiment in which the light pulse is replaced by a material particle of mass m, emitted at speed v with momentum p and kinetic energy K.  The kinetic energy, when deposited in the right end of the box and dissipated as heat, cannot be distinguished from the electromagnetic energy that was deposited in the earlier version.  The total mass transferred from left to right, then, must be m + K/c².  Meanwhile, the box recoiled at speed V = p/M for time t = L/v.  The condition for the center of mass of the box to stay put is MVt = (m + K/c²)L, which gives $p = (m + K/c^2)v$.

As a particle is accelerated from rest, the work done by the net force goes into kinetic energy, so we have dK = F dz = (dp/dt)(v dt) = v dp.  We can eliminate the momentum using $p = (m + K/c^2)v$, then integrate up to find K = m(γ − 1)c².  Alternatively, we can eliminate K and find p = mγv, fully justifying Bell's use of that expression.[5]

## 3 The Quantum Version

Nevertheless, it is a long way from from the flattening of an electron's orbit, computed via purely classical mechanics, to the contraction of macroscopic material objects, the properties of which are governed by quantum mechanics.  Any attempt at a comprehensive grounding of special relativity in single-frame physics must confront the quantum.[6] The aim of this paper is to show how elementary quantum concepts, together with the experimentally observed dependence of energy and momentum on mass and velocity, offer a dynamical route toward the goal.

Quantum theory itself is motivated by experiment.  The photoelectric effect and other phenomena demonstrate that light behaves as though consisting of particles, "photons," for which the energy and momentum are related to frequency and wavelength by E = hf = hc/λ and p = h/λ, where h is Planck's constant.  Atomic line spectra then indicate that the energy of electrons in atoms is quantized.  The existence of discrete states invites interpretation via standing waves, and particle diffraction experiments confirm the existence of matter waves (deBroglie waves) for which the wavelength is again related to the particle momentum by p = h/λ.  These relations, together with the principle of superposition and the uncertainty principle, are the elements of quantum theory being brought to bear in this paper.

The mass-energy relation obtained from Einstein's thought experiment suggests (subject to experimental support from nuclear reactions and mass spectrometry) that energy may be associated directly with the mass of a particle at rest; this "rest energy" is $E_0$ = mc². When traveling at speed v, its total energy is E = $E_0$ + K = mγc² = (p²c²+m²c⁴)^{1/2}.  Then localized material particles can be represented by quantum wavepackets, for which the group velocity is equal to the particle velocity, if the frequency is related to the total energy via E = hf. The frequency associated with a particle at rest, i.e., the rest frequency, is $f_0 = E_0/h = mc^2/h$.

We have, of course, helped ourselves to developments that arrived some time after Einstein's relativity postulates. Nevertheless, those experimental and theoretical advances did not in principle require (even if they were facilitated by) the prior existence of the special theory of relativity. The idea of quantum waves surely would have been conceived and found essential to the understanding of atomic and molecular structure. Nuclear transmutations and high-speed particle interactions would have been observed to conform to the conclusions about mass-energy equivalence and modified expressions for energy and momentum that were already implicit in earlier physics as we have seen. *One could know and apply all these results without suspecting that length contraction and time dilation--much less the covariance of physical laws in relatively moving inertial reference frames—might follow therefrom.* To emphasize: what matters is not the plausibility of any hypothetical sequence of historical events; what matters is that in developing an alternative path to special relativity one does not smuggle Einstein's 1905 postulates into one's premises.

In what follows we examine two concrete examples in which length contraction and time dilation can be derived from the theoretical ingredients assembled above; we find that the frequencies and wavelengths of quantum waves lead naturally to a set of coordinates that obey the Lorentz transformation; and we infer--from the same dynamical considerations--the Lorentz covariance of the equations of classical particle mechanics. On the basis of this dynamical approach, we then suggest a new understanding of the Minkowski geometry of spacetime, one that offers to resolve a recent dispute in philosophy of physics.

## 4 Quantum Dynamics of a Moving Hydrogen Atom

In this section, we use the Heisenberg uncertainty principle to establish a characteristic length scale for a hydrogen atom and show that it contracts by a factor of $1/\gamma$ at high speed. We proceed step by step, considering first a stationary atom, then an atom moving at speeds much less than c, and then the high-speed case. Our heuristic model of the hydrogen atom is a quantum analog of Bell's classical model. In both cases, the essential feature of the analysis is that it is all carried out in a single frame of reference.

For simplicity let us think in terms of purely one-dimensional motion. According to the uncertainty principle, if a particle is restricted to within a distance interval $\Delta z$ there is an associated minimum momentum of order $\hbar/\Delta z$ (where $\hbar = h/2\pi$) and hence a kinetic energy (if $K \ll mc^2$) of order $\hbar^2/2m\Delta z^2$. To be held within the interval $\Delta z$, the potential energy barrier to be overcome must therefore be of the same order. If the restraining force is the attraction of a massive particle of equal and opposite charge q, this barrier is $kq^2/\Delta z$ where k is the Coulomb force constant. Setting these energy expressions equal, we find the particle is confined roughly within a distance

$$\Delta z = \frac{\hbar^2}{2kq^2 m} \ . \tag{1}$$

Using the charge and mass of an electron, this expression gives the correct order of size of a hydrogen atom.

Suppose the atom is moving at velocity v << c. If the electron is restricted to an interval Δz, the corresponding momentum increment Δp (= ℏ/Δz) gives rise to a kinetic energy difference of ΔE = (1/2m) [(p + Δp)² - p²] = vΔp + Δp²/2m where p = mv is the momentum of the electron if traveling at the same velocity as the proton. Unlike in the previous case, this energy cannot be set equal to the potential energy barrier. Rather, the potential energy barrier must be set equal to the net loss of kinetic energy of the electron-proton system as the barrier is overcome, and in the case of the moving atom, the change of kinetic energy of the proton is significant.

If the electron loses momentum Δp while barely escaping to large distance from the proton, the proton must have gained Δp. Then the kinetic energy gained by the proton is (1/2M) [P² - (P - Δp)²] = vΔp - Δp²/2M, where P = Mv is the momentum of the proton (of mass M) at velocity v. Comparing with the energy lost by the electron, and applying M>>m, we see that the net loss in kinetic energy as the electron escapes is given by Δp²/2m = (ℏ/Δz)²/2m. Setting this equal to the potential energy barrier kq²/Δz, we again arrive at Eq. (1).

At high speed, we must use E = mγc² = (p²c² + m²c⁴)^(1/2) for the energy of the electron. A Taylor series to second order tells us that a small momentum change Δp (from p) gives an energy change

$$\Delta E = \frac{pc^2 \Delta p}{(p^2 c^2 + m^2 c^4)^{1/2}} + \frac{1}{2} \left\{ \frac{c^2}{(p^2 c^2 + m^2 c^4)^{1/2}} - \frac{p^2 c^4}{(p^2 c^2 + m^2 c^4)^{3/2}} \right\} \Delta p^2$$

$$= v \Delta p + \frac{\Delta p^2}{2m\gamma^3} \ . \tag{2}$$

For the proton, we replace m by M and p = mγv by P = Mγv, and again, given conservation of momentum, we have ΔP = - Δp. Proceeding as before, when we add the energy increments for the electron and proton, the first-order term in Δp cancels. Other terms are negligible via m<<M. The total ΔE becomes

$$\Delta E = \frac{\Delta p^2}{2m\gamma^3} \ , \tag{3}$$

which corresponds via the uncertainty principle to

$$\Delta E = \frac{\hbar^2}{2m\gamma^3 \Delta z^2} \ . \tag{4}$$

We wish to set this equal to the potential energy barrier.

The expression for the potential barrier is also different at high speeds. In Maxwellian electrodynamics, in the situation we are considering, the electric field of the proton is decreased by a factor of $1/\gamma^2$ along the line of motion, as one can see from equations 1 and 2 in Bell [1] or in standard texts.[7] Integrating up the work that would be done as the separation of the particles increases from $\Delta z$ to infinity, the potential barrier then becomes $kq^2/\gamma^2\Delta z$. By equating with Eq. (4) we arrive at

$$\Delta z = \frac{\hbar^2}{2\gamma k q^2 m} \quad . \tag{5}$$

Comparing with Eq. (1), the factor of $\gamma$ in the denominator represents the Fitzgerald contraction.

## 5 The Free-Particle Wavepacket

The foregoing considerations have appealed explicitly to Maxwellian electrodynamics and have not been truly "relativistic" in the sense of relating observations made in different inertial reference frames. In this section we omit the electrodynamics and demonstrate that the quantum-mechanical wavepacket for a free particle moving at various speeds exhibits both length contraction and time dilation. A particle is localized to a region defined by the constructive interference of a group of quantum waves having a range of momenta. The extent of this region can be measured by observations on identically prepared particles.

For a given quantum particle (such as an electron) at rest, the mass m provides a frequency standard $f_0 = mc^2/h = E_0/h$, which is the frequency of a quantum wave having zero momentum. When moving, the particle has velocity v given by the wave group velocity $v = v_g = df/d(1/\lambda) = dE/dp = pc^2/E$. The frequency associated with the total energy of the moving particle is $f = E/h = \gamma E_0/h = m\gamma c^2/h$, but this is the rate of change of the phase $\phi$ (cycles, not radians) with z held constant: $f = \partial\phi/\partial t$. The rate of change of phase *along the particle trajectory* (as observed in the stationary laboratory, not yet in the frame of reference of the particle itself) is

$$F = \frac{\partial \phi}{\partial t} + v_g \frac{\partial \phi}{\partial z}$$

$$= \frac{E - v_g p}{h} \quad . \tag{6}$$

(It is necessary that $\partial\phi/\partial z$ be equal to $-p/h$, in order that a wavepacket with positive momentum is moving in the $+z$ direction.) Equation (6) reduces to

$$F = \frac{d\phi}{dt} = \frac{m\gamma c^2}{h}\left(1 - \frac{v_g^2}{c^2}\right)$$

$$= f_0/\gamma. \tag{7}$$

*It is at this point that we introduce the idea of comparing observations in a moving frame of reference with observations in the laboratory.* The period of quantum oscillation for, say, an electron at rest establishes a time standard in the laboratory, namely, $T_0 = 1/f_0$. For an observer moving with a particle at velocity v, to whom the particle appears at rest, a period of quantum oscillation along the particle trajectory establishes a time standard in the moving reference frame. What we have demonstrated in Eq. (7) is that according to an observer at rest in the laboratory, the latter time standard (i.e. the period of quantum oscillation $T = 1/F$ along the trajectory of the moving particle) is a factor of γ larger than the former: this is the relativistic time dilation

For the standard of length, we use the length of a wavepacket determined by the dispersion in momentum of the phase waves that comprise it. To demonstrate the principle it will suffice to consider two waves having a (small) momentum difference Δp, interfering to produce a packet of average momentum p traveling at speed $v_g$. The length of the resulting region of constructive interference will be L = h/Δp. We wish to construct comparison wavepackets moving at different speeds, for which the component waves would be the same as determined by an observer moving with the wavepacket. For this purpose, we will take the momentum dispersion Δp in each case to be that which corresponds to the same (small) *fractional* energy shift, hence frequency shift, *as observed in each moving frame*. That is, each Δp is associated with the same fractional change in the number of phase cycles along the path of the particle (the center of the wavepacket). In what follows, we demonstrate that the lengths of wavepackets so chosen depend on the group velocity in accordance with the Lorentz-Fitzgerald contraction formula. The derivation is at once elementary and somewhat tedious.

In Eq. (6), a small energy difference ΔE will be associated with a small momentum difference Δp, but because we are counting phase cycles along the particle path (i.e., the center of the moving wavepacket), $v = v_g$ is held constant in determining

$$h \, \Delta F = \Delta E - v_g \, \Delta p \; . \tag{8}$$

To first order in the Taylor series, from $p^2c^2 + E_0^2 = E^2$ we have

$$\Delta p = \frac{dp}{dE} \Delta E$$

$$= \frac{E}{pc^2} \Delta E$$

$$= \frac{\Delta E}{v_g}, \tag{9}$$

from which follows $\Delta F = 0$, which is just the condition for interfering waves to produce the wavepacket moving at speed $v_g$.

To second order, we note

$$\frac{d^2p}{dE^2} = \frac{d\left(\frac{1}{v}\right)}{dE} \tag{10}$$

which, with $v = c[1 - (E_0/E)^2]^{1/2}$ is

$$\frac{d^2p}{dE^2} = \frac{-E_0}{c^2 \gamma p^2 v} . \tag{11}$$

Eq. (11) permits us to write $\Delta p$ in Eq. (8) to second order (in $\Delta E$), with the result

$$h \, \Delta F = \frac{E_0 \Delta E^2}{2c^2 \gamma p^2} . \tag{12}$$

The fractional shift in frequency is

$$\frac{\Delta F}{F} = \frac{h \, \Delta F}{E_0/\gamma}$$

$$= \frac{1}{2c^2}\left(\frac{\Delta E}{p}\right)^2 \tag{13}$$

which is constant by our construction (that is, our comparison wavepackets have the same fractional dispersion in F). This means that when comparing packets of different speeds we must have $\Delta E$ proportional to p, but then $\Delta p = E\Delta E/pc^2$ is proportional to E and thereby to $\gamma$. Then the wavepacket length L, which is proportional to $1/\Delta p$, is proportional to $1/\gamma$. This result we recognize as the Fitzgerald contraction of a wavepacket due to its motion at speed v.

A wavepacket will also disperse with time, because the superposition of many phase waves implies a corresponding spread of group velocities $\Delta v_g$, determined by the spread in energy $\Delta E$ and momentum $\Delta p$. The relevant dispersion timescale is given by $\tau_D = L/(\Delta v_g)$. A bit of algebra applied to the foregoing results demonstrates that this timescale is proportional to $\gamma$ of the wavepacket, providing us with another example of time dilation. We mention this example in case the earlier use of the frequency associated with rest mass should meet with objection on the grounds that that frequency is not directly observable.

## 6 Quantum Coordinates

So far, it has been demonstrated that length and time standards based on the wavelength and frequency of quantum waves will be subject to the same length contraction and time dilation as the rods and clocks in Einstein's special relativity. It is only a short step further to show that a system of coordinates constructed from these standards will transform between inertial frames via the familiar Lorentz transformation. The key lies in recognizing that for a particle at rest, the constant-phase wavefronts are constant-time wavefronts. That is (in one dimension), if one draws the wavefronts on a graph of t versus z (a "spacetime diagram"), horizontal lines will be lines of constant phase. For a particle in motion with velocity v, the lines of constant phase will be such that $\Delta\phi = E\Delta t - p\Delta z = 0$, or $\Delta t = (p/E)\Delta z = v\Delta z/c^2$. But these lines of constant phase must serve as the lines of constant time in the reference frame of an observer moving with the particle in question.

Therefore, taking t' and z' as the time and space coordinates in the moving reference frame, we see that for $\Delta t' = 0$ we have $\Delta t = v\Delta z/c^2$. For the case of $\Delta z' = 0$ we have already obtained $\Delta t = \gamma\Delta t'$. In addition, since the primed reference frame is moving with velocity v, the condition $\Delta z' = 0$ necessarily implies $\Delta z = v\Delta t$. Furthermore, the Fitzgerald length contraction implies that for $\Delta t = 0$ we will have $\Delta z = \Delta z'/\gamma$ (the length of the wavepacket in the laboratory is the distance between its front and back at the same instant). The linear transformation satisfying all these cases is

$$\Delta z' = \gamma(\Delta z - v\, \Delta t) \qquad (14a)$$
$$\Delta t' = \gamma(\Delta t - v\, \Delta z/c^2) \qquad . \qquad (14b)$$

Algebraically solving for $\Delta z$ and $\Delta t$ in terms of $\Delta z'$ and $\Delta t'$, we find the inverse transformation

$$\Delta z = \gamma(\Delta z' + v\, \Delta t') \qquad (15a)$$
$$\Delta t = \gamma(\Delta t' + v\, \Delta z'/c^2) \qquad . \qquad (15b)$$

Thus, the grid of space and time coordinates in the laboratory and in the moving frame of reference are related by what we recognize as the Lorentz transformation, the symmetry of which offers no distinction between these frames. Our laboratory time and length standards will be observed, from the moving frame of reference, to be subject to the same dilation and contraction that the moving standards are, as observed in the laboratory.

## 7 Lorentz Covariance of Physics: How Far Have We Come?

As we have seen, the phase generated along the path of a particle in time dt is $d\phi = f_0\, dt/\gamma$ so that we have

$$d\phi^2 = f_0^2\, dt^2/\gamma^2 = f_0^2\, dt^2\, (1 - v^2/c^2) = f_0^2\, (dt^2 - dz^2/c^2). \qquad (16)$$

Now the elapsed phase of a wave on different paths determines whether interference is constructive or destructive, a matter on which observers must agree. Elapsed phase is simply a count of wave cycles. Eq. (16) therefore defines an invariant. Normalized to the rest frequency $f_0$, it is just the invariant interval of Minkowski space. In textbooks, the Lorentz transformation is derived from the invariance of this quantity.

Because $dt/\gamma$ is invariant, $p = m\gamma \frac{dz}{dt}$ transforms as dz and $E = m\gamma c^2$ transforms as dt. Therefore, E and p transform together as what we call a four-vector under a Lorentz transformation. The quantities p and t transform to $p' = m\gamma' \frac{dz'}{dt'}$ and $E' = m\gamma' c^2$. Furthermore, the elapsed phase

$$d\phi = \frac{\partial \phi}{\partial t} dt + \frac{\partial \phi}{\partial x} dz = (Edt - pdz)/h$$

transforms to

$$d\phi = \frac{\partial \phi}{\partial t'} dt' + \frac{\partial \phi}{\partial x'} dz' = (E'dt' - p'dz')/h$$

and $hd\phi$ is just the increment in action for a particle in classical mechanics. In order that elapsed quantum phase be an invariant for all particle mechanics, the same transformation properties must apply to the energy and momentum associated with each of the forces of nature individually, hence also with the equations involving their source terms. As an example, the electromagnetic contribution to the energy of a charged particle is expressed through the electric ("scalar") potential, while the contribution to momentum comes from the vector potential. When a system is put into motion, these potentials must change in the same way as energy and momentum generally. That is why, in special relativity, the potentials comprise a four-vector that transforms, as the energy-momentum four-vector does, via the Lorentz transformation. The bottom line is that the laws of classical particle mechanics must take the same form in reference frames for which the coordinates are related by the Lorentz transformation: that is, they are Lorentz covariant.

  What about rods and clocks? Is it possible, for example, that macroscopic rods do not undergo length contraction, even if free hydrogen atoms and free-particle wavepackets do? What about properties such as spin and processes such as particle creation? If the dynamical construction of special relativity is to encompass all of physics, it will take a longer story, but perhaps the outline is emerging. A metal rod is a stable configuration of atoms. The locations of the atoms are determined by a maximum square amplitude of a wavefunction in a configuration space. If we can represent wavefunctions using Fourier superpositions of pure waves whose wavelengths and frequencies are related to momentum and energy, then, since we know how momentum and energy transform, we should be able to infer what the wavefunction looks like—and hence, what the rod looks like—in other reference frames. Can this rude sketch be developed into a complete picture? It does not seem quite so dauntingly ambitious an undertaking as had been feared (see Brown[2005], p. 148, n. 55).

  We alluded earlier to a parallel with thermodynamics, for which Einstein saw the foundations as supplied by statistical mechanical models such as the kinetic theory of gases. Like the second law of thermodynamics, Einstein's relativity postulates serve, putatively, as a constraint on all future physics. Efforts to prove the second law, with its arrow of time, from statistical mechanics have generated a vast and enlightening literature going back to the nineteenth century. Nevertheless, the subject remains contentious. With that comparison in mind, if we have not arrived at a rigorous and comprehensive dynamical grounding of special relativity, we have nevertheless traveled some distance along a promising path toward the goal.

# 8 The Nature of Minkowski Spacetime

We are now in position to address a recent dispute in philosophy of physics regarding the nature of Minkowski spacetime. Let a meterstick be gently accelerated from rest to 0.8c in our (very large) laboratory. Our instruments measure its length and find it to be only 60 centimeters. What caused it to shrink? We can of course infer the result from the relativity postulates, via the Lorentz transformation: from Eq. (14a), given $\Delta z' = 100 \, cm$, with $\Delta t = 0$ we find $\Delta z = \Delta z'/\gamma = 60 \, cm$. But the fact that we can calculate the correct answer by applying the relativity postulates to a comparison of two rest frames does not tell us how an object's acceleration to high speed through our laboratory caused its length to contract. The laboratory has not been accelerated. Its measuring devices remain unchanged. Whatever happened, happened to the meterstick.

In the (dominant) spacetime realist view, length contraction is due to the independently existing Minkowski geometry of spacetime, to which physical systems, through the dynamical laws governing their behavior, must conform. Norton ([2008], p. 823) gives a clear summary of this view:

> (1) There exists a four-dimensional spacetime that can be coordinatized by a set of standard coordinates (x, y, z, t), related by the Lorentz transformation.
>
> (2) The spatiotemporal interval s between events (x, y, z, t) and (X, Y, Z, T) along a straight line connecting them is a property of the spacetime, independent of the matter it contains, and is given by [with c = 1]
>
> $$s^2 = (t - T)^2 - (x - X)^2 - (y - Y)^2 - (z - Z)^2.$$
>
> When $s^2 > 0$, the interval s corresponds to times elapsed on an ideal clock; when $s^2 < 0$, the interval s corresponds to spatial distances measured by ideal rods (both employed in the standard way).
>
> (3) Material clocks and rods measure these times and distances because the laws of the matter theories that govern them are adapted to the independent geometry of this spacetime.

Thus, Minkowski geometry comes first. It constrains the dynamics and, through the dynamics, the properties of rods and clocks.

A competing view, advocated by spacetime "constructivists," puts the dynamics first. Dynamical laws govern the properties of physical systems. An essential fact about these laws is that they are Lorentz covariant. Lorentz covariance is what ensures that space and time intervals measured by rods and clocks in relative motion are related in the manner that spacetime realists would attribute to an independently existing geometry. The view is expressed in Brown's endorsement ([2005], p. 133) of "the idea that Lorentz contraction is the result of a structural property of the forces responsible for the microstructure of matter...The appropriate [geometrical] structure is Minkowski geometry *precisely because* the laws of physics of the non-gravitational interactions are Lorentz covariant." [Emphasis in the original.]

The quantum dynamical approach developed in this paper supports the spacetime constructivists' contention that dynamics comes first. What spacetime realists attribute to the *geometry* of spacetime actually arises from Lorentz covariance of the dynamical laws governing the properties of physical systems such as rods and clocks. Nevertheless, the dynamics itself is to be found in spacetime, specifically in the refractive properties of spacetime for the propagation of quantum waves. These properties would not go away in the absence of physical particles, any more than the refractive properties of glass go away when the light is turned off. The spacetime realist view is thus vindicated in that spacetime itself constrains physical laws and physical systems. The quantum dynamical approach explains why the results lend themselves so readily to geometric interpretation, by identifying the invariant interval of special relativity with the elapsed phase of a particle propagating freely between two events (divided by the rest frequency). This phase is the measure of the proper time along the particle path, for $s^2 > 0$.

For $s^2 < 0$, identifying s directly with distance measured by a meterstick requires ignoring a factor of $i = (-1)^{1/2}$. Identifying s with phase suggests a different possibility. Quantum mechanically, what is the significance of an imaginary phase? Consider a spatial wave of the form $e^{ikx}$. If the wavevector k is imaginary, the result is not a wave but an exponentially decaying real quantity. Now, in quantum terms, an imaginary k would mean an imaginary momentum and a negative kinetic energy. This is what happens where the wavefunction penetrates into a classically forbidden region. Traveling faster than light is classically forbidden, but nevertheless it is not quantum mechanically entirely forbidden. The dynamical approach indicates that for $v > c$ the wavefunction undergoes an exponential decay, and the amount of the decay becomes a measure of distance.

## 9 Summary

This paper has introduced a quantum dynamical approach to special relativity. The aim is to ground the relativity postulates in physics that can be independently established in a single inertial frame of reference. Thereby, relativity would be furnished with the "elementary foundations" that Einstein himself found it to lack, in much the way that statistical mechanics has been employed in an effort to ground the second law of thermodynamics. The program is far from complete. Nevertheless, we have demonstrated that  a) "relativistic" particle dynamics can be established independently of the relativity postulates,  b) a rudimentary one-dimensional hydrogen atom undergoes length contraction,  c) a quantum wavepacket exhibits both contraction and time dilation, d) a system of coordinates based on the frequencies and wavelengths of quantum waves obeys the Lorentz transformation, and  e) the laws of classical particle mechanics must be Lorentz covariant. The quantum dynamical approach offers to resolve the dispute between spacetime realists and constructivists regarding the nature of Minkowski spacetime, by agreeing with the constructivists on the primacy of dynamics while locating the dynamics in the independently existing properties of spacetime.

**Footnotes**

[1] Accessible historical accounts are found in Brown [2005], chapters 1, 4 and Darrigol [2000], chapters 8, 9.

[2] Lecturing in 1906, Lorentz mildly insinuated that Einstein had jumped to the right answer without sufficient foundation: "[Einstein's] results concerning electromagnetic and optical phenomena...agree in the main with those which we have obtained in the preceding pages, the chief difference being that Einstein simply postulates what we have deduced...." Reproduced in Lorentz [1916], pp. 229-230.

[3] On thermodynamics and Einstein's principle versus constructive theory distinction, see Brown [2005], chapter 5 and Klein [1967]. For a convenient entry point to recent philosophical debates, see Frisch [2011].

[4] Bell's aim is both pedagogical and foundational. Our interest here is foundational. See also Brown [2005] chapter 7, Miller [2010], and Shanahan [2014].

[5] French ([1968], pp. 20-22) also obtains the relativistic expressions for momentum and energy prior to taking up the relativity postulates, but only by speculating that one can combine $E = mc^2$ with the Newtonian expression $p = mv$ to obtain $E = c^2 p/v$. The modified thought experiment described in this paper gives the result a secure basis.

[6] Swann [1941] is cited in this regard by both Bell [1982] and Brown [2005].

[7] See, for example, Becker [1964], pp. 269-271. Note especially Eq. 64.18.